%% file: main.tex
\documentclass[conference]{IEEEtran}
\IEEEoverridecommandlockouts
\usepackage{balance}
\usepackage{cite}
\usepackage{amsmath,amssymb,amsfonts}
\usepackage{algorithmic}
\usepackage{graphicx}
\usepackage{textcomp}
\usepackage{xcolor}
\usepackage{tikz}
\usepackage{booktabs}
\def\BibTeX{{\rm B\kern-.05em{\sc i\kern-.025em b}\kern-.08em
    T\kern-.1667em\lower.7ex\hbox{E}\kern-.125emX}}
\begin{document}

	\title{
    Measurement-Driven O-RAN Diagnostics with Tail Latency and Scheduler Indicators
\vspace{-0.3cm}
    }

\author{
\IEEEauthorblockN{ 
Theofanis P. Raptis\IEEEauthorrefmark{1}
Weronika Maria Bachan\IEEEauthorrefmark{2},
Roberto Verdone\IEEEauthorrefmark{3}
}
\IEEEauthorblockA{
\IEEEauthorblockA{\IEEEauthorrefmark{1}Institute of Informatics and Telematics, National Research Council, Pisa, Italy}
\IEEEauthorblockA{\IEEEauthorrefmark{2}WiLab, National Inter-University Consortium for Telecommunications,
Bologna, Italy}
\IEEEauthorblockA{\IEEEauthorrefmark{3}Department of Electrical, Electronic and Information Engineering, ``Guglielmo Marconi'', University of Bologna, Italy
}}
Email: theofanis.raptis@iit.cnr.it, weronikamaria.bachan@wilab.cnit.it, roberto.verdone@unibo.it
\vspace{-0.5cm}
}

\maketitle

\begin{tikzpicture}[remember picture,overlay]
\node[anchor=south,yshift=10pt] at (current page.south) {\fbox{\parbox{\dimexpr\textwidth-\fboxsep-\fboxrule\relax}{
  \footnotesize{
     \copyright 2026 IEEE. Personal use of this material is permitted.  Permission from IEEE must be obtained for all other uses, in any current or future media, including reprinting/republishing this material for advertising or promotional purposes, creating new collective works, for resale or redistribution to servers or lists, or reuse of any copyrighted component of this work in other works.
  }
}}};
\end{tikzpicture}

\begin{abstract}
We investigate cross-layer performance diagnostics for an O-RAN instance by jointly analyzing application-level latency and radio-layer behavior from a real measurement campaign. Measurements were conducted at multiple link distances (2, 6 and 11 meters) using two representative UE configurations (a commercial smartphone and a modem-based device), under both static conditions and a controlled dynamic obstruction scenario. Rather than relying on averages, the study adopts tail-focused latency characterization (e.g., 95th percentile and exceedance probabilities) and connects it to scheduler- and link-adaptation indicators (e.g., block error behavior, modulation/coding selection and signal quality). The results reveal (i) UE-dependent differences that primarily manifest in the latency tail, (ii) systematic scaling of tail latency with distance and payload and (iii) cases where radio-layer dynamics are detectable even when end-to-end latency appears stable, motivating the need for cross-layer evidence. Distinct from much of the existing literature (often centered on throughput, simulated setups, or single-layer KPIs) this work contributes a measurement-driven methodology for interpretable O-RAN diagnostics and proposes lightweight, window-based ``degradation flags'' that combine tail latency and radio indicators to support practical monitoring and troubleshooting.
\end{abstract}



\section{Introduction}
O-RAN architectures are accelerating the transition from monolithic cellular infrastructures to disaggregated, software-driven RAN deployments, enabling openness, programmability and multi-vendor innovation \cite{11124199}. At the same time, this increased modularity complicates performance assurance: user-perceived quality (e.g., interactive latency) emerges from a stack of interacting components spanning the application, transport and radio layers. As a result, diagnosing performance degradations in operational O-RAN settings remains challenging \cite{10024837}, especially when symptoms are intermittent and dominated by rare events rather than average behavior.

A recurring gap in existing empirical studies \cite{10621427} is the limited integration of \emph{tail-aware} application metrics with radio-layer evidence. Many measurement papers focus on throughput or average latency \cite{9039732}, which may overlook sporadic but impactful ``stall'' events that dominate user experience. Conversely, RAN-facing monitoring often relies on scheduler and link-adaptation indicators that are not directly connected to end-to-end behavior. Bridging these views is particularly relevant for industrial O-RAN deployments, where practical troubleshooting requires interpretable cross-layer signals that can support root-cause triage without assuming access to proprietary internal components \cite{8764545}.

This paper takes a measurement-driven step toward such cross-layer diagnostics by jointly analyzing end-to-end ICMP round-trip times and gNB-side scheduler/link-adaptation indicators collected across multiple distances, UE types and propagation conditions (including a controlled dynamic obstruction scenario). The focus is not on proposing a new learning model or a new protocol, but on building a transferable \emph{methodology} and a compact set of analysis artifacts that turn heterogeneous logs into interpretable evidence about (i) latency tails, (ii) radio reliability/adaptation dynamics and (iii) their coupling over time. The main contributions of this paper are:
\begin{itemize}
  \item We measure and report robust and tail-aware latency indicators (median, p95, exceedance probabilities) across distances, packet sizes and UE types, highlighting how differences primarily manifest in the tail rather than in central tendency.
  \item We complement end-to-end observations with gNB-side indicators related to reliability and link adaptation (e.g., BLER, MCS, SNR), showing cases where radio-layer dynamics are visible even when latency appears stable.
  \item We introduce a window-based analysis that aligns latency tails with scheduler excursions and demonstrates simple ``degradation flags'' suitable for practical monitoring and troubleshooting workflows.
\end{itemize}

The remainder of the paper is organized as follows. In Section~\ref{sec:related}, we present some related works. Section~\ref{sec:dataset} describes the measurement scenarios and dataset structure. Section~\ref{sec:methodology} details the processing pipeline and the metrics used. Section~\ref{sec:results} presents the exploratory results, including latency-tail behavior, scheduler-side dynamics and cross-layer coupling. Section~\ref{sec:limitations} discusses limitations and future potential and Section~\ref{sec:conclusions} concludes the paper.

\section{Related Work}
\label{sec:related}

Open and disaggregated RAN architectures (including O-RAN) increase observability and programmability, but they also make performance diagnosis harder because impairments can originate at different layers and interfaces. Recent work has started to exploit this increased observability for monitoring: Permal \emph{et al.} propose continuous latency monitoring using Open RAN interfaces, aiming at always-on performance visibility rather than one-off trials~\cite{permal2024latmon}. Complementarily, large-scale Open RAN testbeds (e.g., OpenRAN@Brasil) report deployment-centric performance observations and operational lessons, highlighting the importance of actionable KPIs under realistic conditions~\cite{medeiros2025openranbr}.

A parallel line of work focuses on \emph{where latency is introduced} inside the RAN stack. Hamici \emph{et al.} dissect 5G NR latency across interacting layers and trade-offs, showing that end-to-end latency emerges from the joint behavior of scheduling, retransmissions and protocol timing (rather than a single bottleneck)~\cite{hamici2025dissect}. At a more system-profiling level, 5GPerf benchmarks open-source 5G RAN components under different architectural choices, helping identify implementation-dependent latency and throughput sensitivities~\cite{wei2022fivegperf}.

Finally, SDR-based experimental C-RAN/O-RAN platforms have been used to analyze latency budgets and practical constraints introduced by disaggregation and real-time processing. Laskos \emph{et al.} study latency in SDR-based experimental C-RAN/O-RAN systems, providing evidence on how architectural and processing choices affect latency at runtime~\cite{laskos2025oranlat}.

\textbf{Gap and novelty.} While prior work addresses monitoring, component-level profiling, or broad latency decomposition, fewer studies provide a \emph{tail-aware, cross-layer} diagnostic view that jointly (i) quantifies end-to-end latency tails and (ii) links them to scheduler-side reliability/adaptation indicators within the same measurement campaign. This paper contributes such a view by combining tail-focused latency indicators with radio/scheduler metrics and by introducing lightweight, windowed degradation flags that are practical for deployment diagnostics even when the underlying dataset cannot be publicly released.

\section{Dataset Overview} \label{sec:dataset}
The received dataset comprises:
\begin{itemize}
  \item \textbf{Ping (ICMP) latency logs}: end-to-end round-trip times for packet sizes of 30~B and 1000~B, sampled every 0.2~s, for multiple distances (2~m, 6~m, 11~m) and scenarios (static, spatial averaging and a dynamic ``people moving'' test).
  \item \textbf{gNB statistics logs}: ``fullstats'' CSV traces containing MAC/PHY indicators (e.g., BLER/MCS and related parameters) to support interpretation.
\end{itemize}
For 11~m, only smartphone data are available, consistent with the measurement notes. A dynamic test includes 15~minutes LOS followed by 15~minutes with human movement between gNB and UE. For the 2~m and 6~m campaigns, measurements were complemented with spatial averaging of small-scale fading to improve reliability, as performed during data collection. Each distance-specific ping campaign lasted 30~minutes per position. In addition, a 60~minute static acquisition at 6~m (without spatial averaging) and a 30~minute static LOS-to-dynamic-obstruction test (15~min+15~min) were performed.

\section{Processing and Methodology} \label{sec:methodology}

\subsection{Data consolidation}
The raw measurement logs were first organized and harmonized into a consistent, analysis-ready representation. In particular, the latency traces collected via ICMP ping were parsed into a unified dataset where each observation is associated with its experimental context (UE type, distance, packet size and scenario/run). In parallel, the gNB ``fullstats'' logs were ingested into a structured format that preserves the relevant PHY/MAC indicators and identifiers (including per-UE identifiers where available), enabling subsequent cross-layer interpretation. 

To support both rapid inspection and systematic reporting, summary statistics were computed at the level of individual runs (e.g., sample counts and descriptive latency statistics) and an additional time-windowed representation was derived by aggregating measurements over short, fixed-duration intervals with overlap. This windowed view is particularly useful for analyzing temporal dynamics, identifying transient anomalies and facilitating correlation with radio-layer events on comparable time scales.

\subsection{Metrics}

The primary observable in this paper is the end-to-end round-trip time measured via ICMP ping, which provides an application-facing view of network responsiveness. To characterize latency in a way that is informative for both typical behavior and rare degradations, we report robust statistics such as the median (representative ``typical'' latency) and the 95th percentile (capturing upper-tail behavior experienced by a non-negligible fraction of packets). In addition, we quantify tail indicators through outlier rates above predefined thresholds (e.g., occurrences of unusually large delays), which are particularly relevant for identifying sporadic stalls that may be masked by averages. When interpreting BLER-related fields, the definitions and computation follow the OAI logging conventions as documented in the OAI MAC documentation.

Figures are employed with two complementary purposes. First, they serve as sanity checks to validate that each run exhibits plausible timing, continuity and sample density. Second, they provide baseline visual evidence of how latency distributions and temporal dynamics differ across UE types, packet sizes, link distances and static versus dynamic conditions. Where appropriate, these application-level observations can be interpreted jointly with gNB-side indicators in follow-on analyses to link end-to-end effects to underlying PHY/MAC adaptations.


\section{Exploratory Results} \label{sec:results}
\subsection{Latency distributions at 6 m (baseline)}
\begin{figure}[t!]
  \centering
  \includegraphics[width=0.5\linewidth]{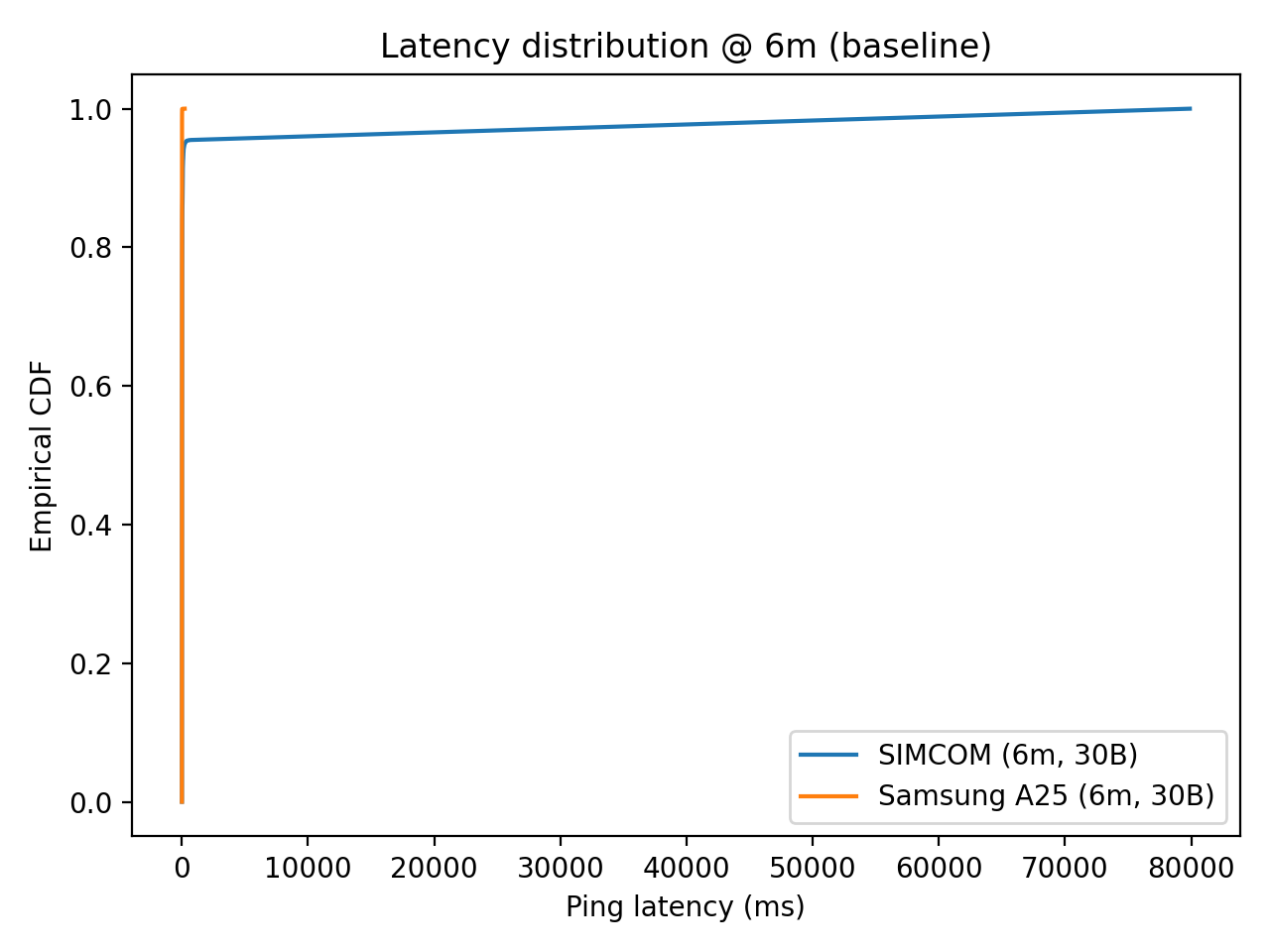}
  \caption{Latency CDF at 6~m, 30~B baseline: comparison between smartphone and modem-based UE.}
  \label{fig:cdf}
\end{figure}

Figure~\ref{fig:cdf} reports a baseline latency CDF at 6~m for 30~B packets, comparing UE types. The CDF highlights differences in distribution tails, which are important for understanding rare but impactful latency spikes. In the considered setting, the two UE types exhibit markedly different latency behavior. The smartphone curve rises steeply and reaches values close to 1 at very low latencies, indicating that the vast majority of ping samples are concentrated in a narrow region (i.e., low dispersion and limited jitter). In contrast, the modem-based UE shows a much more gradual approach to 1: although a substantial fraction of samples still fall in the low-latency region, the curve develops a  long tail extending to very large values. This tail implies that a non-negligible subset of packets experiences rare but extreme delays (multi-second “stall” events), which are effectively invisible when reporting only averages or medians. Overall, the CDF highlights that, under the same distance and packet-size conditions, the smartphone provides a more stable end-to-end latency profile, while the modem-based UE exhibits significantly higher variability driven by infrequent but impactful outliers. This observation motivates complementing application-level latency analysis with gNB-side indicators (e.g., BLER/MCS dynamics) to investigate whether tail events correlate with transient radio-layer degradations, scheduling effects, or UE-specific protocol/stack behavior.

\subsection{Packet-size and UE comparison (6 m baseline)}
\begin{figure}[t!]
  \centering
  \includegraphics[width=0.5\linewidth]{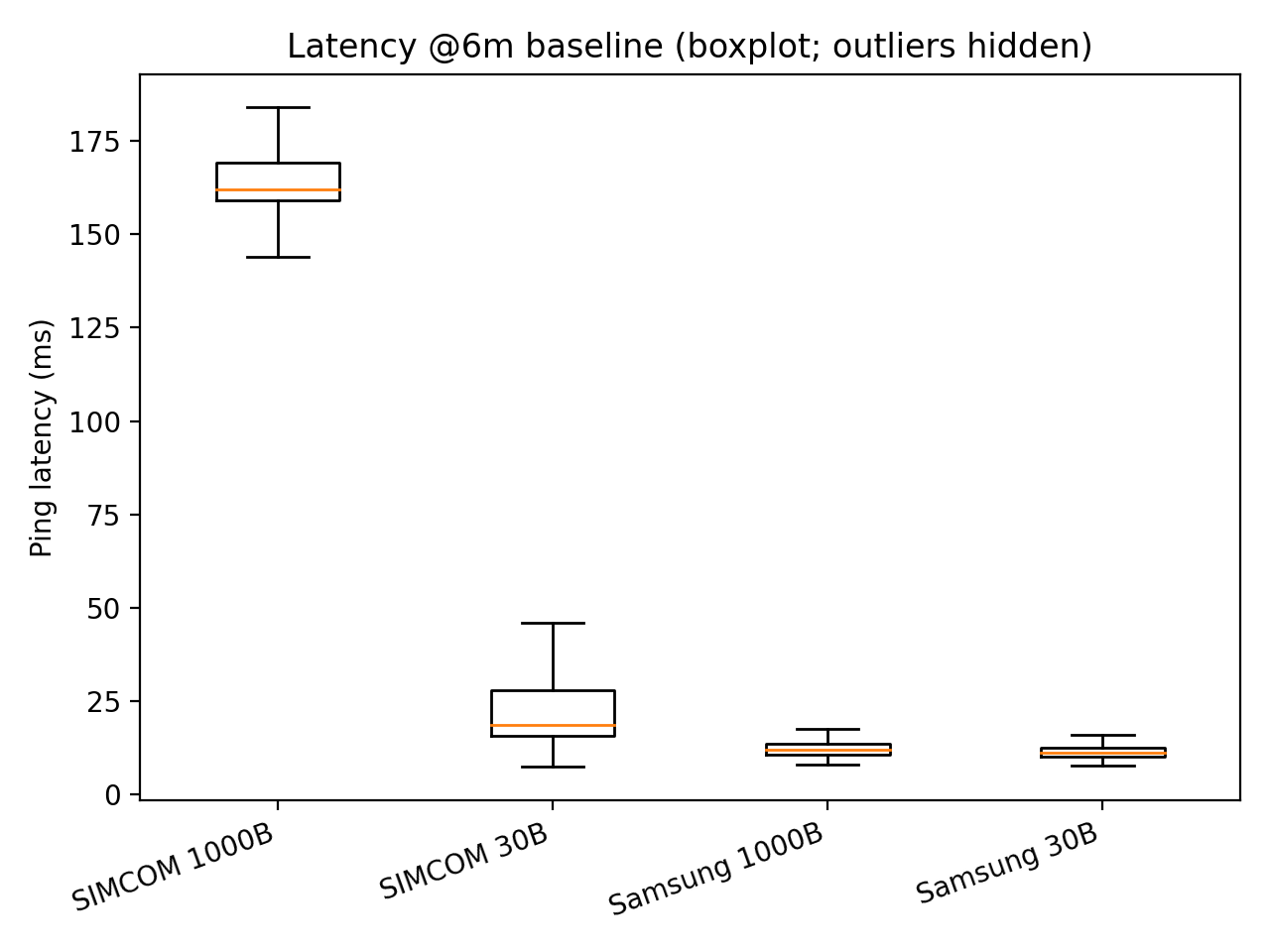}
  \caption{Latency boxplot at 6~m baseline: 30~B vs 1000~B, smartphone vs modem-based UE.}
  \label{fig:box}
\end{figure}

Figure~\ref{fig:box} summarizes latency variability across packet sizes and UE types at 6~m. This view complements the CDF by providing a compact comparison of central tendency and dispersion. Specifically, Figure~\ref{fig:box} provides a compact comparison of end-to-end latency across the two packet sizes (30~B and 1000~B) and the two UE types at 6~m by visualizing medians, interquartile ranges and outliers. For the smartphone, the boxes are narrow and the medians remain in the low-millisecond region for both packet sizes, confirming a generally stable latency profile with limited dispersion; occasional outliers may occur, but they are sparse and remain relatively bounded. In contrast, the modem-based UE exhibits substantially wider boxes and longer whiskers, indicating markedly higher variability even when the median latency is comparable to the smartphone case. The difference becomes particularly evident when moving from 30~B to 1000~B packets, where the modem-based UE shows an upward shift and increased spread, consistent with stronger sensitivity to payload size and/or buffering and scheduling effects along the stack. Overall, the boxplot complements the CDF by making the dispersion and tail behavior immediately visible: it reinforces that the smartphone behaves consistently across packet sizes, while the modem-based device is characterized by significantly larger jitter and a higher incidence of extreme latency events under the same link distance.

\subsection{Distance scaling and tail exceedance (baseline)}
\begin{figure}[t!]
  \centering
  \includegraphics[width=0.5\linewidth]{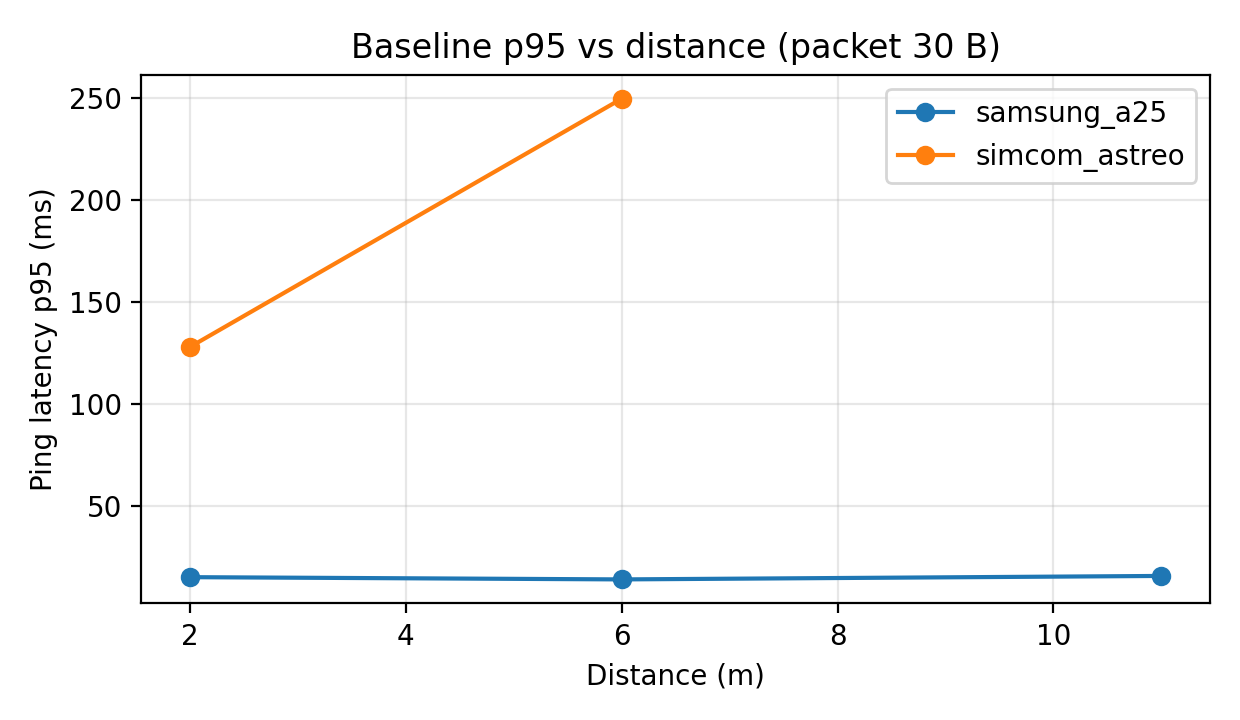}
  \caption{Baseline 95th-percentile latency vs distance for 30~B packets (per UE where available).}
  \label{fig:p95dist30}
\end{figure}

\begin{figure}[t!]
  \centering
  \includegraphics[width=0.5\linewidth]{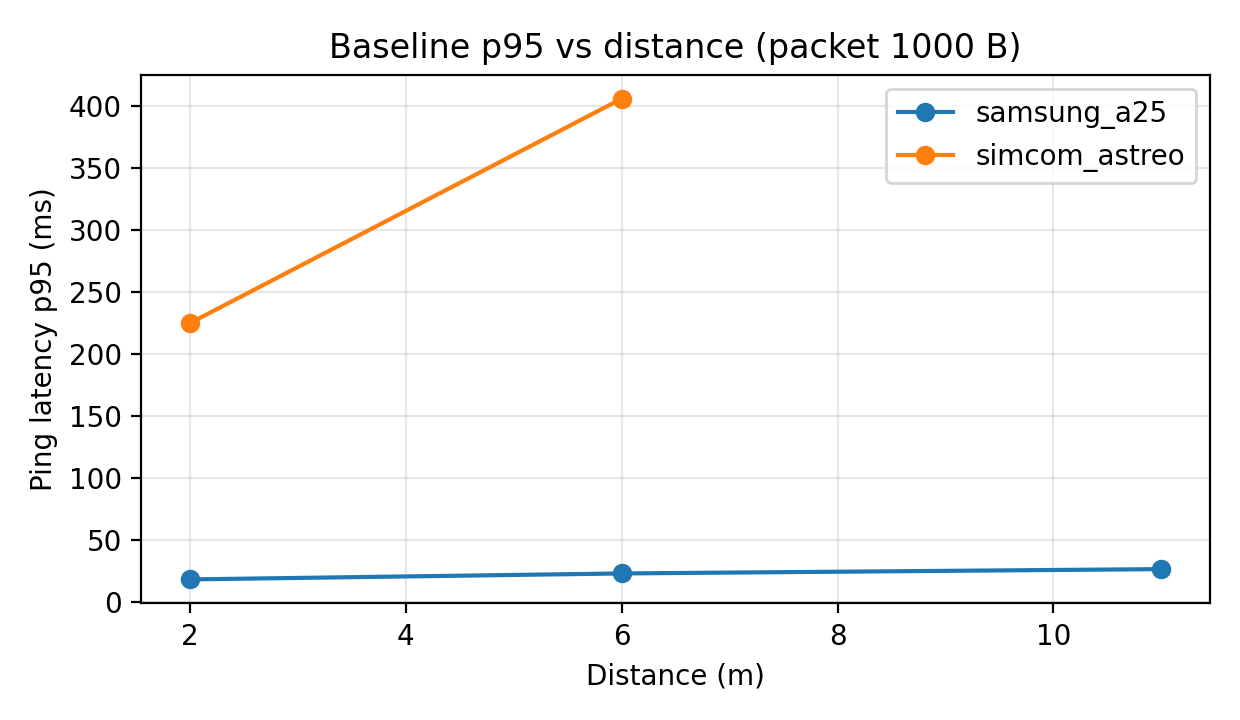}
  \caption{Baseline 95th-percentile latency vs distance for 1000~B packets (per UE where available).}
  \label{fig:p95dist1000}
\end{figure}

While Fig.~\ref{fig:cdf}-\ref{fig:box} focus on 6~m, the dataset also includes 2~m and 11~m acquisitions (smartphone only at 11~m). To provide a concise scaling view, Fig.~\ref{fig:p95dist30} and Fig.~\ref{fig:p95dist1000} report the 95th-percentile latency across distances for 30~B and 1000~B packets, respectively (baseline runs). Tail latency increases with distance and larger packets amplify upper-tail sensitivity. Importantly, UE type remains a key factor: even at the same distance, the modem-based UE shows substantially higher tail variability than the smartphone, consistent with the long-tail behavior observed at 6~m.

\begin{figure}[t!]
  \centering
  \includegraphics[width=0.5\linewidth]{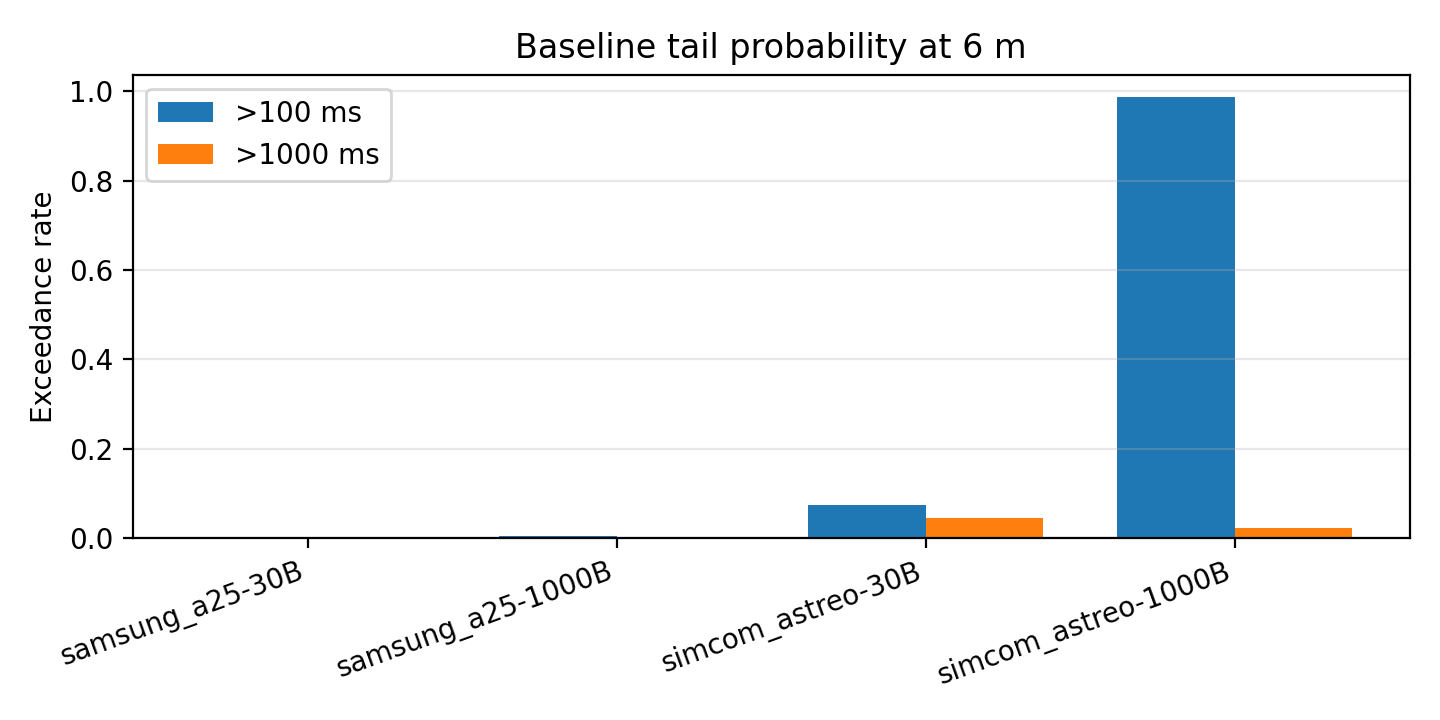}
  \caption{Baseline exceedance probabilities at 6~m (by UE and packet size) for 100~ms and 1~s thresholds.}
  \label{fig:exceed}
\end{figure}

Tail behavior is further summarized in Fig.~\ref{fig:exceed}, which reports exceedance probabilities for two practical thresholds (100~ms and 1~s) at 6~m. This complements percentile-based reporting: even when p95 values remain low, rare ``stall'' events above 1~s can dominate perceived responsiveness and motivate diagnostics beyond averages.

For completeness, Table~\ref{tab:ks_ue} reports a two-sample Kolmogorov-Smirnov (KS) test on the baseline 6~m/30~B latency distributions, confirming that the UE-specific latency distributions differ beyond noise.

\input{tables/ks_ue_comparison.tex}

\subsection{Dynamic ``people moving'' test (smartphone, 6 m)}
\begin{figure}[t!]
  \centering
  \includegraphics[width=0.5\linewidth]{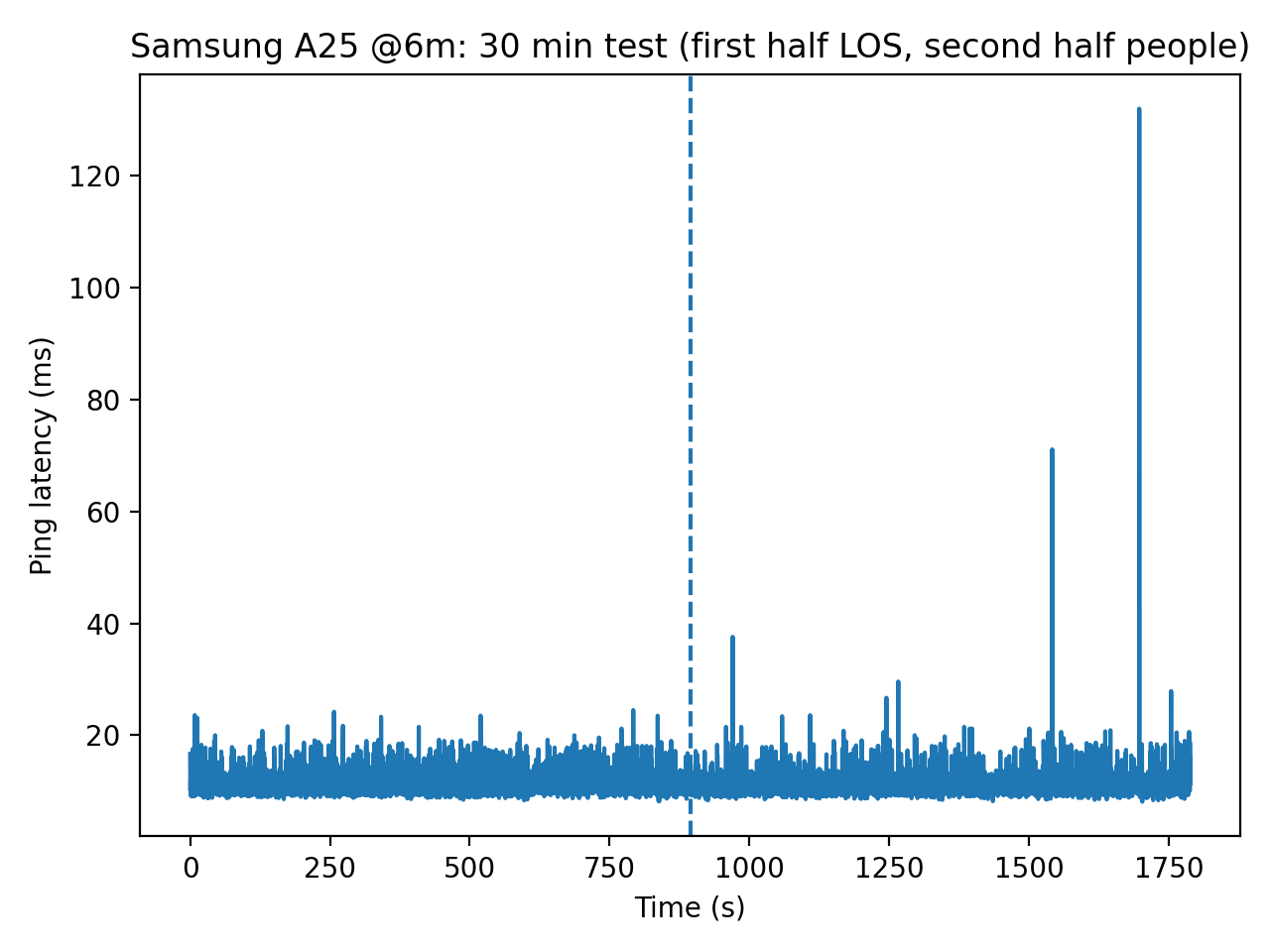}
  \caption{Latency time series for the smartphone in the dynamic obstruction test (6~m).}
  \label{fig:ts}
\end{figure}

Figure~\ref{fig:ts} reports the time evolution of ping latency for the smartphone during the dynamic obstruction experiment, in which the first 15 minutes correspond to a clear line-of-sight (LOS) condition and the subsequent 15 minutes include human movement between gNB and UE, introducing time-varying blockage. The time-series view is useful to verify (i) the continuity and duration of the trace, (ii) the baseline level of latency and its short-term fluctuations and (iii) the presence (or absence) of visible regime changes when transitioning from LOS to dynamic obstacles. In this run, latency appears largely stable over the full 30-minute period, with only sporadic spikes that do not form a persistent shift in the baseline. This outcome is informative in itself: it suggests that, at least for the smartphone under the tested distance and traffic pattern, application-level round-trip time can remain robust even when the propagation environment becomes dynamically obstructed, possibly due to adequate link margin, fast adaptation at lower layers and/or buffering that masks transient radio impairments. At the same time, the absence of a clear latency step does not imply that the radio channel is unaffected; rather, it motivates a joint, time-aligned inspection of gNB-side indicators (e.g., BLER, MCS, retransmissions and scheduling statistics) to detect subtle degradations or compensatory adaptations that may not directly translate into higher end-to-end ping latency but are still relevant for reliability, capacity and energy efficiency.

\subsection{Dynamic obstruction: phase-wise view}
\begin{figure}[t!]
  \centering
  \includegraphics[width=0.5\linewidth]{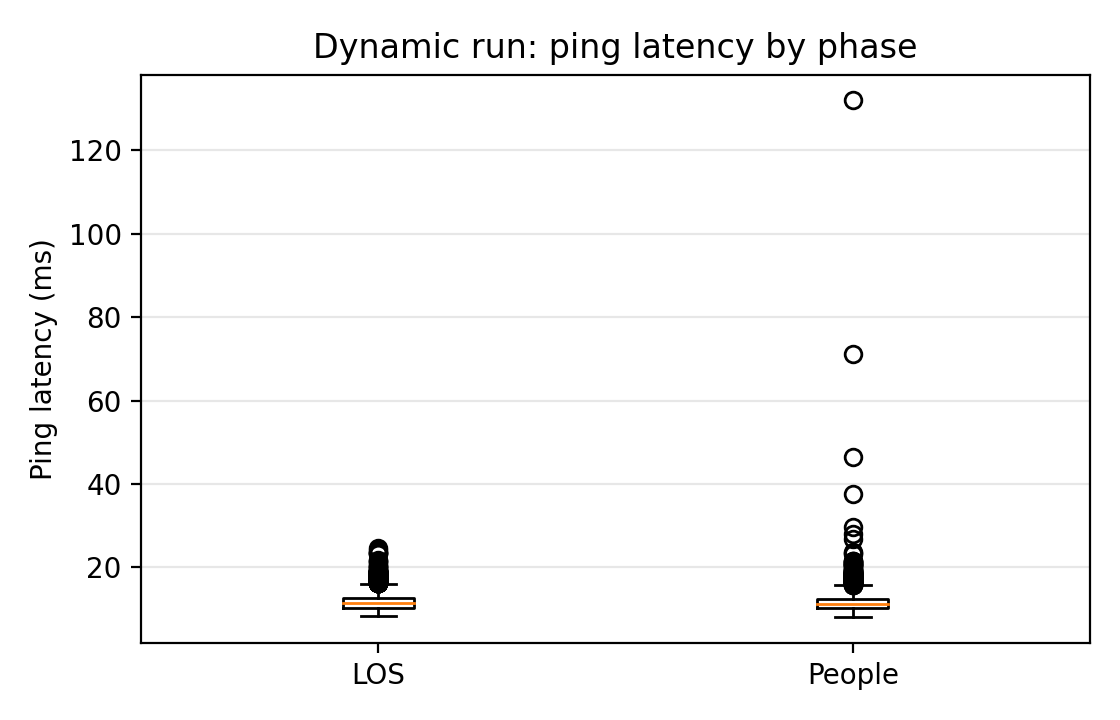}
  \caption{Dynamic run (smartphone, 6~m/30~B): latency distribution by phase (LOS vs people).}
  \label{fig:dyn_ping_box}
\end{figure}

The dynamic campaign is useful to test sensitivity to controlled environmental changes. Beyond the full-run time series (Fig.~\ref{fig:ts}), Fig.~\ref{fig:dyn_ping_box} compares the first 15~minutes (LOS) versus the second 15~minutes (with human movement) using distribution summaries. In this run, latency medians and p95 values remain close across phases, indicating that end-to-end ICMP latency alone may not provide a strong separation between LOS and obstructed conditions for the smartphone at 6~m.

\begin{figure}[t!]
  \centering
  \includegraphics[width=0.5\linewidth]{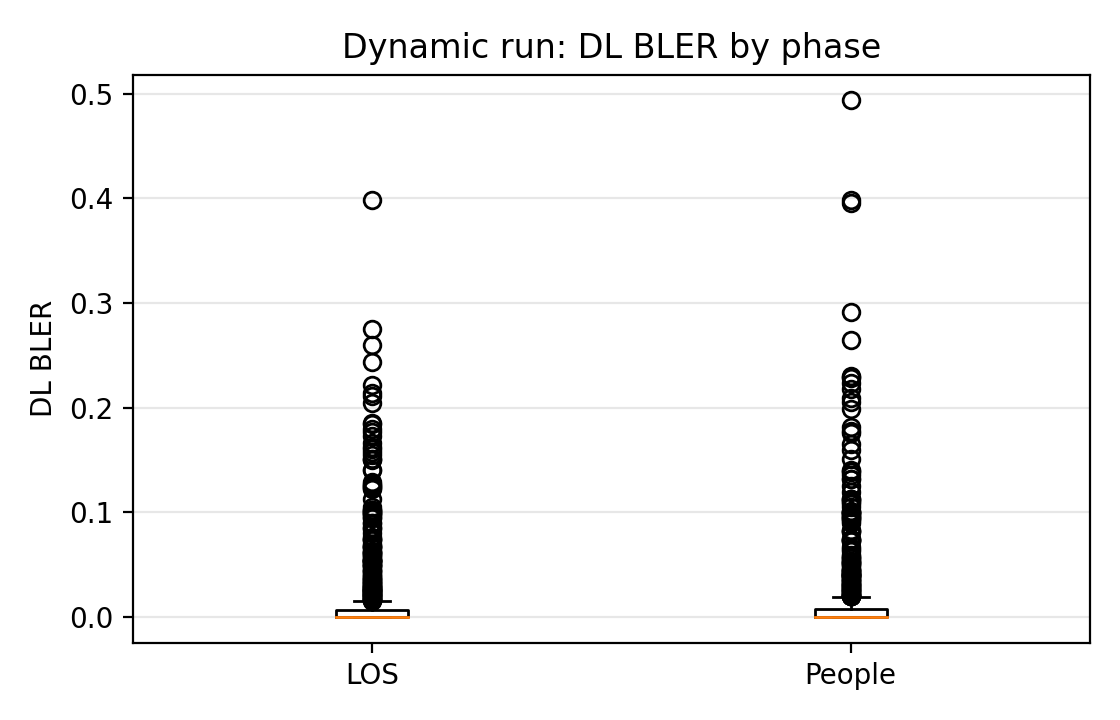}
  \caption{Dynamic run (smartphone, 6~m/30~B): DL BLER distribution by phase (dominant UE instance).}
  \label{fig:dyn_bler_box}
\end{figure}

Scheduler-side indicators, however, can still exhibit phase-dependent variability. Fig.~\ref{fig:dyn_bler_box} reports the corresponding DL BLER distributions by phase and Table~\ref{tab:dynamic_phase} summarizes latency-tail and BLER-tail indicators side by side. Even when the ping-level signal is stable, this view can expose radio-layer stress that is compensated by link adaptation.

\input{tables/dynamic_phase_summary.tex}

\subsection{Windowed cross-layer coupling and lightweight diagnostics}
\begin{figure}[t!]
  \centering
  \includegraphics[width=0.5\linewidth]{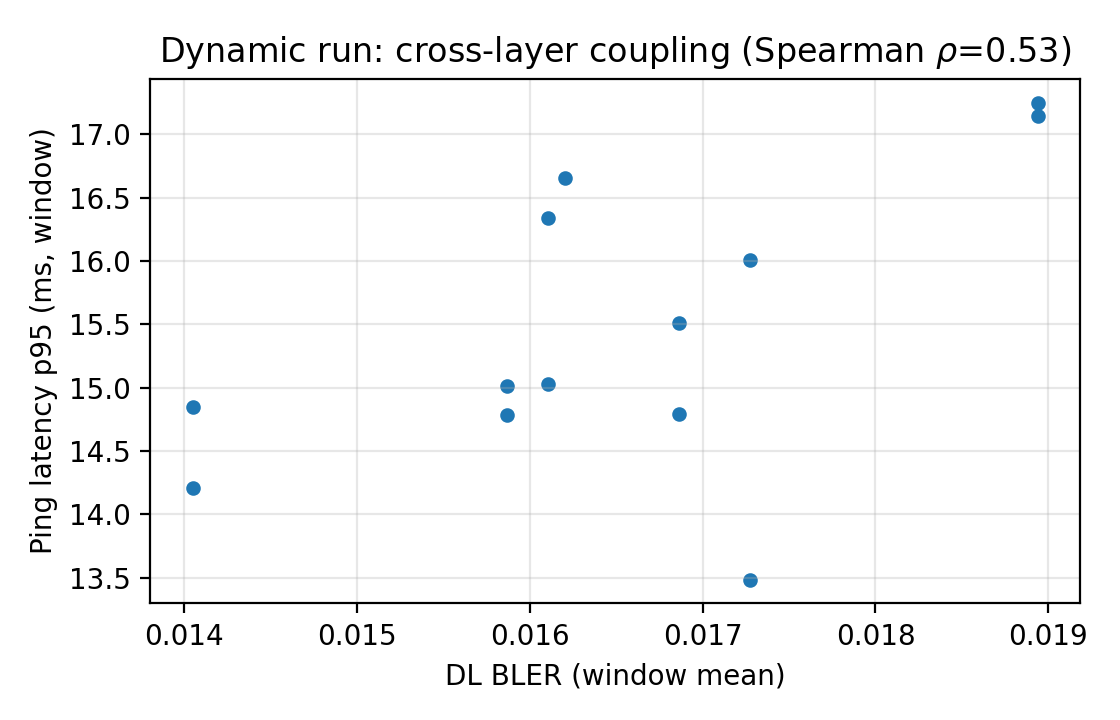}
  \caption{Dynamic run (smartphone, 6~m/30~B): windowed scatter between latency p95 and BLER mean.}
  \label{fig:cross_scatter}
\end{figure}

To move from descriptive plots toward diagnostics, we adopt a windowed representation (10~s windows, 5~s stride) and join ping windows with corresponding scheduler windows. Fig.~\ref{fig:cross_scatter} illustrates the relationship between windowed latency tails and BLER and Table~\ref{tab:crosslayer_corr} quantifies this effect with Spearman correlation coefficients.

\input{tables/crosslayer_corr_table.tex}

\begin{figure}[t!]
  \centering
  \includegraphics[width=0.5\linewidth]{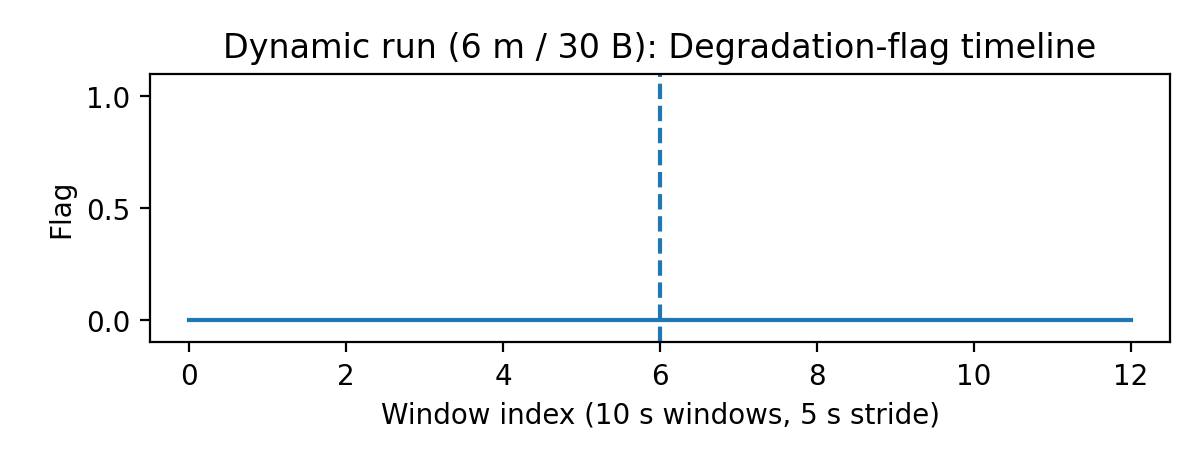}
  \caption{Dynamic run (smartphone, 6~m/30~B): example windowed degradation-flag timeline (10~s windows, 5~s stride).}
  \label{fig:flag_ts}
\end{figure}

Finally, to demonstrate an operationally relevant outcome, we define a lightweight ``degradation flag'' per window that triggers when both (i) the latency tail exceeds a threshold and (ii) a scheduler-side excursion is present (e.g., elevated BLER). Fig.~\ref{fig:flag_ts} visualizes the flag timeline for the dynamic run and Table~\ref{tab:flags} reports illustrative flag rates across scenarios.

\input{tables/degradation_flag_rates.tex}

\subsection{Scheduler-side metrics (gNB fullstats) at 6 m}
In addition to end-to-end ping latency, the dataset includes gNB ``fullstats'' logs that expose PHY/MAC indicators produced by the OAI MAC scheduler (e.g., DL/UL BLER, MCS, SNR, RSRP, retransmission counters). These indicators are useful because they provide a direct view of link adaptation and channel quality, which may or may not surface as noticeable changes in ICMP round-trip time.

It is important to note that, in OAI, the reported BLER is not a generic PHY BLER estimate: it is computed from HARQ behavior over a recent window (e.g., the ratio of first-round retransmissions to total transmissions over the last window) and it is explicitly used by the scheduler logic to adapt the selected MCS via upper/lower BLER targets. Therefore, BLER and MCS should be interpreted jointly: BLER excursions tend to trigger downward MCS adjustments, while persistently low BLER allows the scheduler to increase MCS.

Moreover, the exported fullstats CSVs may contain multiple RNTIs (e.g., due to reconnects or multiple UE instances observed during logging). For concise reporting, the following results focus on the \emph{dominant} UE instance per run (i.e., the RNTI with the largest number of records), which provides a stable reference for cross-run comparison.

\begin{table*}[t!]
\centering
\small
\setlength{\tabcolsep}{4pt}
\caption{Scheduler-side summary at 6~m and 30~B (dominant UE instance per run).}\label{tab:sched_summary}
\begin{tabular}{l c c c c c}
\hline
Scenario & N & $\mathrm{BLER}_{DL}$ (med.) & $\mathrm{BLER}_{DL}$ (p95) & $\mathrm{MCS}_{DL}$ (med.) & SNR (med., dB) \\
\hline
Baseline (avg) & 1592 & 0.053 & 0.326 & 6 & 27.5 \\
Dynamic (people) & 1448 & 0.000 & 0.100 & 9 & 28.5 \\
Static (no avg, 60 min) & 4539 & 0.000 & 0.000 & 9 & 30.5 \\
\hline
\end{tabular}
\vspace{-0.6cm}
\end{table*}

\begin{figure}[t!]
  \centering
  \includegraphics[width=0.5\linewidth]{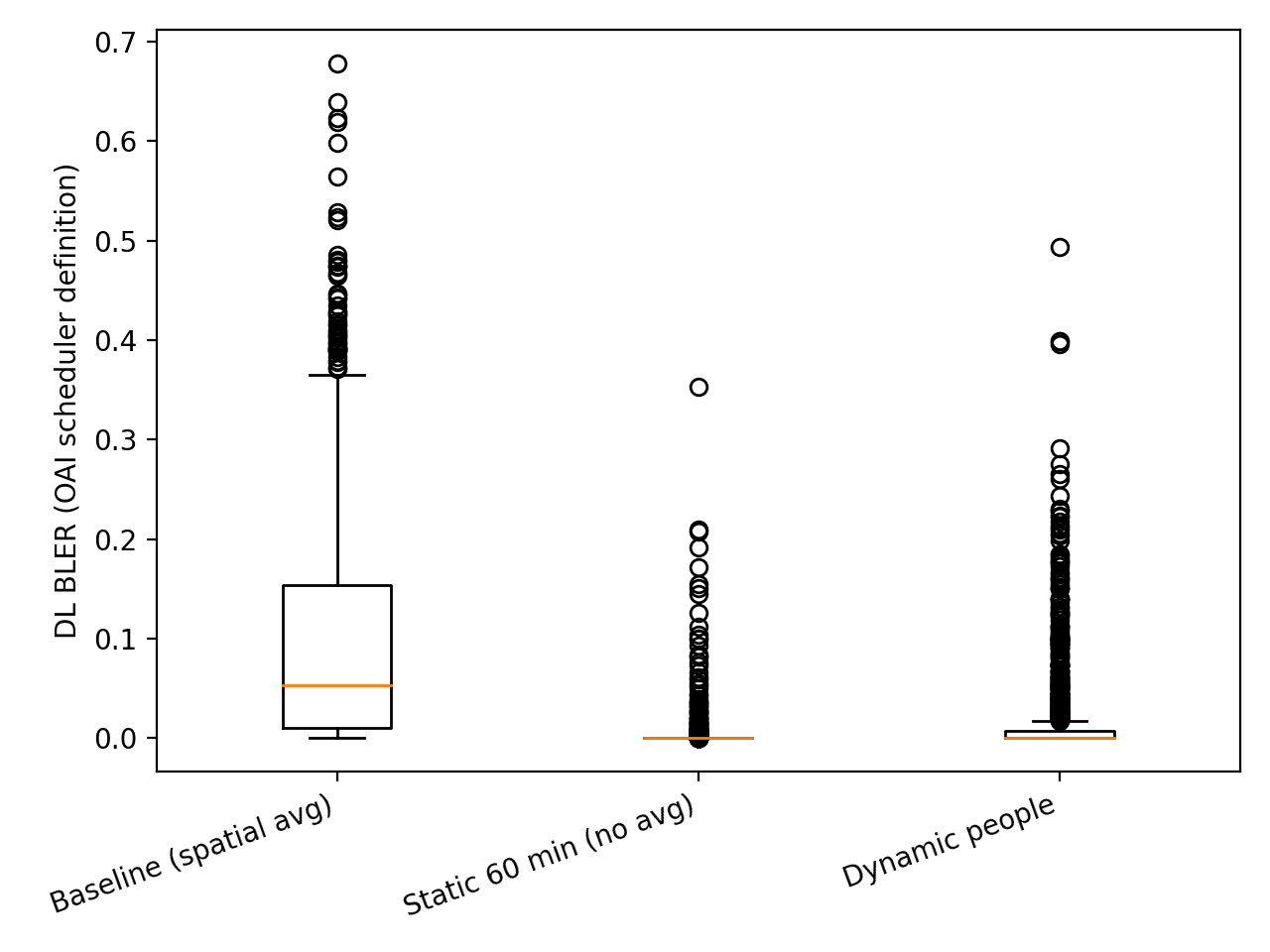}
  \caption{DL BLER distribution at 6~m and 30~B across baseline (with spatial averaging in the measurement protocol), static no-averaging run and dynamic ``people moving'' run.}
  \label{fig:sched_bler_box}
\end{figure}

Figure~\ref{fig:sched_bler_box} provides direct scheduler-side evidence of how radio reliability behaves across scenarios at 6~m. In particular, it visualizes not only central tendency but also the presence of excursions (outliers) in the BLER signal that can trigger link adaptation. This view is complementary to the ping-based latency plots: it can reveal radio-layer stress even when end-to-end latency does not show a clear regime shift.

\begin{figure}[t!]
  \centering
  \includegraphics[width=0.5\linewidth]{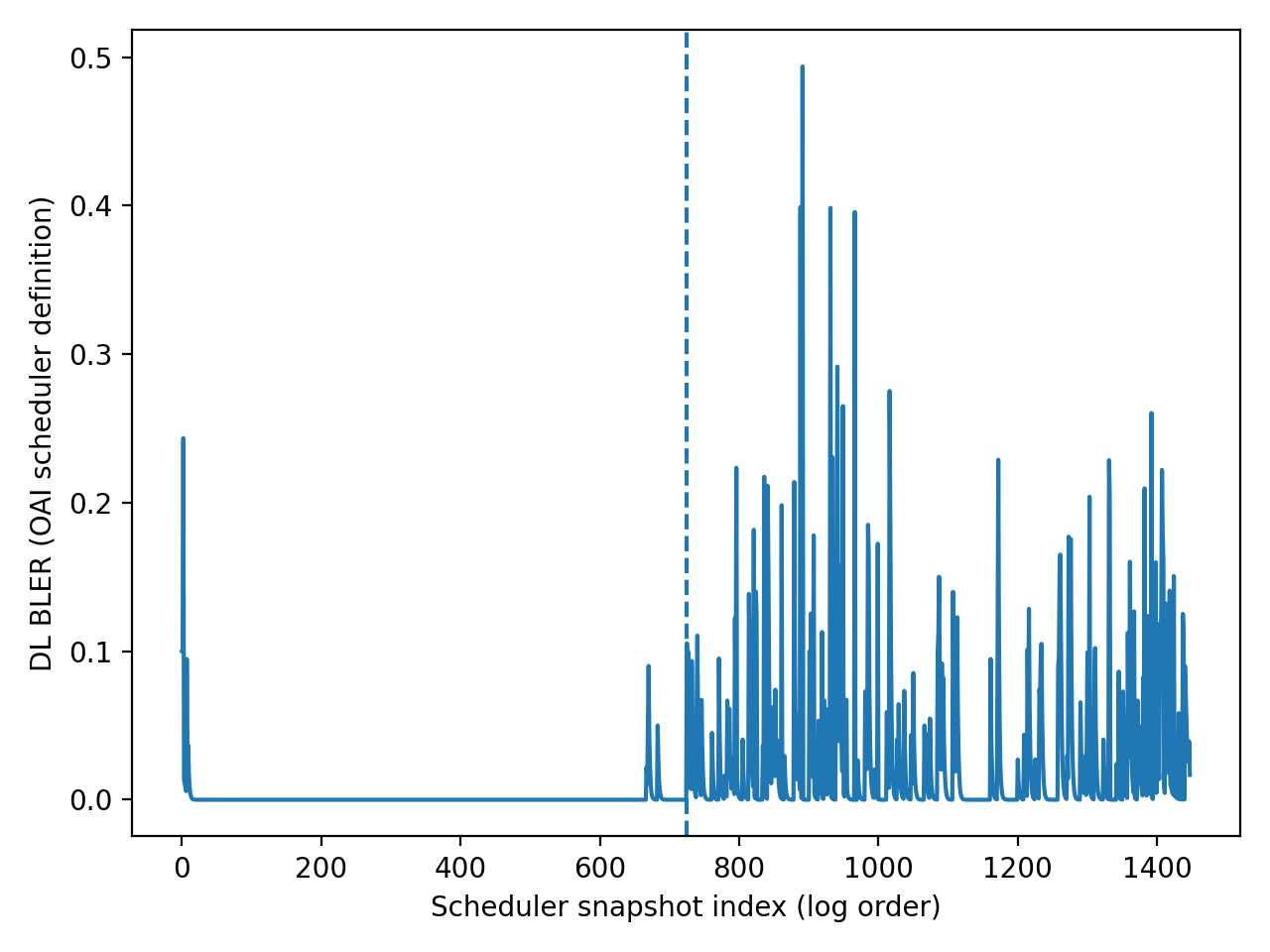}
  \caption{DL BLER over scheduler snapshot index for the dynamic obstruction run at 6~m (dominant UE instance). The dashed marker indicates the midpoint of the run (nominally the LOS-to-obstruction transition per measurement protocol).}
  \label{fig:sched_bler_ts}
\end{figure}

Figure~\ref{fig:sched_bler_ts} shows the short-term dynamics of DL BLER during the dynamic obstruction test. While the ping latency time series for the smartphone may appear largely stable, scheduler-side BLER can still exhibit localized excursions that are quickly compensated through MAC/PHY adaptations (e.g., retransmissions and MCS adjustments). This motivates the next-step analysis of time-aligned correlation between latency tails and scheduler indicators to distinguish radio-driven events from stack- or buffering-driven artifacts.

\section{Limitations and Future Potential} \label{sec:limitations}
The present dataset provides broad experimental coverage, yet it also exhibits natural constraints typical of real testbed campaigns. Coverage differs across distances and UE types (e.g., measurements at 11~m are available only for the smartphone) and the current results are primarily derived from ICMP ping, which captures application-level round-trip time but may mask transient radio-layer adaptations (e.g., changes in MCS, HARQ retransmissions, or scheduling decisions) that do not necessarily translate into sustained latency increases. Consequently, the findings reported here should be interpreted as a baseline assessment of end-to-end behavior, rather than a complete diagnosis of underlying PHY/MAC dynamics.

Beyond these limitations, the dataset offers strong potential for deeper, cross-layer analysis and for deriving actionable indicators for monitoring and troubleshooting O-RAN deployments. In particular, the following directions can extend the current work from descriptive characterization to explainable and operational insights:
\begin{enumerate}
  \item \textbf{Cross-layer time alignment}: establish a consistent time base between ping traces and gNB statistics to enable direct correlation between application-level latency and radio/MAC events.
  \item \textbf{Explainable coupling analysis}: quantify how BLER/MCS/RSRP (and related parameters, when available) co-vary with latency dynamics across distances and scenarios, with emphasis on identifying signatures of dynamic blockage and short-term channel fluctuations.
  \item \textbf{Operational indicators and alerts}: define lightweight ``degradation flags'' that combine radio-layer indicators (e.g., BLER excursions or abrupt MCS shifts) with application-layer tail behavior (e.g., increased probability of high-latency outliers), with the longer-term goal of supporting automated monitoring, anomaly detection and root-cause triage in O-RAN systems.
\end{enumerate}

\balance
\section{Conclusions} \label{sec:conclusions}

This paper provides a demonstration of a first use of an O-RAN measurement dataset and establishes baselines at both the application level (ICMP latency) and scheduler level (gNB fullstats). In the 6 m latency-only analysis, different UE configurations show markedly different tail behavior: even with similar medians, some runs include rare but consequential spikes that averages miss. This supports using tail-aware metrics (e.g., 95th percentile and outlier rates) to reflect perceived responsiveness.
Scheduler indicators improve interpretability. In the 6 m / 30 B subset, gNB fullstats reveal scenario-dependent DL BLER behavior and corresponding MCS selection, consistent with OAI’s scheduler logic where BLER excursions trigger MCS adaptation. The static long run provides a stable reference (low BLER dispersion, sustained high MCS), while baseline and dynamic runs enable stress-testing under protocol changes (spatial averaging) and time-varying obstruction. Crucially, ping step changes alone do not prove radio-channel changes; scheduler metrics can expose radio-layer dynamics masked end-to-end by buffering and fast adaptation.

Overall, the data suggest that (i) application-level differences are mainly driven by tail events and (ii) gNB scheduler metrics are essential to attribute tails to radio-layer reliability/adaptation versus higher-layer effects. The key follow-on is a time-aligned cross-layer analysis linking ping tails to BLER/MCS/SNR excursions, yielding lightweight diagnostic “degradation flags” for monitoring O-RAN deployments. Distance-scaling further shows tail latency increases with distance and larger packets and exceedance analysis (e.g., $>$1 s stalls) separates UE behaviors even when medians match.

\section*{Acknowledgements}

This work was supported by the European Union - Next Generation EU under the Italian National Recovery and Resilience Plan (NRRP), Mission 4, Component 2, Investment 1.3, CUP B53C22003970001, partnership on ``Telecommunications of the Future'' (PE00000001 - program ``RESTART'').

\bibliographystyle{IEEEtran}
\bibliography{references}

\end{document}

%% file: tables/ks_ue_comparison.tex
\begin{table}[t!]
\centering
\small
\caption{Two-sample KS test on ping latency at 6~m baseline (samsung\_a25 vs simcom\_astreo).}\label{tab:ks_ue}
\setlength{\tabcolsep}{4pt}
\begin{tabular}{l c c c c c c}
\toprule
Packet & $n_1$ & $n_2$ & KS stat. & $p$-value & $p95_1$ (ms) & $p95_2$ (ms) \\
\midrule
30 B & 8945 & 8957 & 0.888 & 0.00e+00 & 14.2 & 249.6 \\
1000 B & 8945 & 8630 & 0.985 & 0.00e+00 & 23.2 & 406.0 \\
\bottomrule
\end{tabular}
\end{table}

%% file: tables/dynamic_phase_summary.tex
\begin{table}[t!]
\centering
\small
\caption{Dynamic run (6~m/30~B): phase-wise comparison (first 15 min LOS vs second 15 min with people).}\label{tab:dynamic_phase}
\setlength{\tabcolsep}{4pt}
\begin{tabular}{l c c c}
\toprule
Phase & Lat. p95 (ms) & $P(>100\,ms)$ & BLER p95 \\
\midrule
LOS & 15.7 & 0.000 & 0.100 \\
People & 15.3 & 0.000 & 0.100 \\
\bottomrule
\end{tabular}
\end{table}

%% file: tables/crosslayer_corr_table.tex
\begin{table}[t!]
\centering
\small
\caption{Windowed correlation between latency tail and scheduler indicators (dynamic run, 6~m/30~B).}\label{tab:crosslayer_corr}
\setlength{\tabcolsep}{4pt}
\begin{tabular}{l c}
\toprule
Metric & Value \\
\midrule
Spearman $\rho$(p95 latency, BLER mean) & 0.53 \\
Spearman $\rho$(p95 latency, MCS median) & N/A \\
Windows (joined) & 13 \\
\bottomrule
\end{tabular}
\end{table}

%% file: tables/degradation_flag_rates.tex
\begin{table}[t!]
\centering
\small
\caption{Illustrative degradation-flag rates (10~s windows, 5~s stride) for the smartphone at 6~m/30~B.}\label{tab:flags}
\setlength{\tabcolsep}{4pt}
\begin{tabular}{l c c}
\toprule
Scenario & Windows & Flag rate \\
\midrule
baseline & 13 & 0.077 \\
dynamic people & 13 & 0.000 \\
static 1h & 13 & 0.000 \\
\bottomrule
\end{tabular}
\end{table}